\documentclass[twocolumn,superscriptaddress]{revtex4}

\usepackage{color}
\usepackage{mathtools}
\usepackage{graphicx}
\usepackage{SIunits}

\newcommand{\dograph}[3]{
	\begin{figure}
	\includegraphics[#2]{#1}
	\caption{\label{fig:#1}#3}
	\end{figure}
}

\begin{document}

\title{{Quasiparticle tunneling as a probe of Josephson junction barrier and capacitor material in superconducting qubits}}

\author{C.~Kurter}
\thanks{corresponding author}
\email[email address:]{cihan.kurter@ibm.com}
\affiliation{IBM Quantum, IBM T.J.~Watson Research Center, Yorktown Heights, NY 10598, USA}

\author{C.E.~Murray}
\affiliation{IBM Quantum, IBM T.J.~Watson Research Center, Yorktown Heights, NY 10598, USA}

\author{R.T.~Gordon}
\affiliation{IBM Quantum, IBM T.J.~Watson Research Center, Yorktown Heights, NY 10598, USA}

\author{B.B.~Wymore}
\affiliation{IBM Quantum, IBM T.J.~Watson Research Center, Yorktown Heights, NY 10598, USA}

\author{M.~Sandberg}
\affiliation{IBM Quantum, IBM T.J.~Watson Research Center, Yorktown Heights, NY 10598, USA}

\author{R.~M.~Shelby}
\affiliation{IBM Quantum, IBM Almaden Research Center, San Jose, CA 95120, USA}

\author{A.~Eddins}
\affiliation{IBM Quantum, IBM Almaden Research Center, San Jose, CA 95120, USA}

\author{V.~P.~Adiga}
\affiliation{IBM Quantum, IBM T.J.~Watson Research Center, Yorktown Heights, NY 10598, USA}

\author{A.~D.~K.~Finck}
\affiliation{IBM Quantum, IBM T.J.~Watson Research Center, Yorktown Heights, NY 10598, USA}

\author{E.~Rivera}
\affiliation{IBM Quantum, IBM T.J.~Watson Research Center, Yorktown Heights, NY 10598, USA}

\author{A.~A.~Stabile}
\affiliation{IBM Quantum, IBM T.J.~Watson Research Center, Yorktown Heights, NY 10598, USA}

\author{B.~Trimm}
\affiliation{IBM Quantum, IBM T.J.~Watson Research Center, Yorktown Heights, NY 10598, USA}

\author{B.~Wacaser}
\affiliation{IBM Quantum, IBM T.J.~Watson Research Center, Yorktown Heights, NY 10598, USA}

\author{K.~Balakrishnan}
\affiliation{IBM Quantum, IBM T.J.~Watson Research Center, Yorktown Heights, NY 10598, USA}

\author{A.~Pyzyna}
\affiliation{IBM Quantum, IBM T.J.~Watson Research Center, Yorktown Heights, NY 10598, USA}

\author{J.~Sleight}
\affiliation{IBM Quantum, IBM T.J.~Watson Research Center, Yorktown Heights, NY 10598, USA}

\author{M.~Steffen}
\affiliation{IBM Quantum, IBM T.J.~Watson Research Center, Yorktown Heights, NY 10598, USA}

\author{K.~Rodbell}
\affiliation{IBM Quantum, IBM T.J.~Watson Research Center, Yorktown Heights, NY 10598, USA}

\date{\today}

\begin{abstract}

Non-equilibrium quasiparticles are possible sources for decoherence in superconducting qubits because they can lead to energy decay or dephasing upon tunneling across Josephson junctions (JJs). Here, we investigate the impact of the intrinsic properties of two-dimensional transmon qubits on quasiparticle tunneling (QPT) and discuss how we can use quasiparticle dynamics to gain critical information about the quality of JJ barrier. We find the tunneling rate of the non-equilibrium quasiparticles to be sensitive to the choice of the shunting capacitor material and their geometry in qubits. In some devices, we observe an anomalous temperature dependence of the QPT rate below 100 mK that deviates from a constant background associated with non-equilibrium quasiparticles. We speculate that this behavior is caused by high transmission sites/defects within the oxide barriers of the JJs, leading to spatially localized subgap states. We model this by assuming that such defects generate regions with a smaller effective gap. Our results present a unique \emph{in situ} characterization tool to assess the uniformity of tunnel barriers in qubit junctions and shed light on how quasiparticles can interact with various elements of the qubit circuit. 


\end{abstract}

\maketitle

\section*{Introduction}

There has been a tremendous amount of work recently undertaken towards building a scalable fault-tolerant quantum computer based on superconducting qubits~\cite{DiCarlo:2009vl, Devoret1169}. Some architectures seek to utilize quantum error correction protocols~\cite{Gambetta:2017ue, Corcoles:2015wh} to mitigate errors caused by non-ideal behavior of physical qubits. However, the coherence times of such qubits still need to be enhanced to meet the requirements for the error correction threshold~\cite{PhysRevLett.107.240501, MIT2020}. One possible mechanism that can limit the qubit coherence times is the presence of non equilibrium quasiparticles~\cite{PhysRevLett.103.097002, PhysRevLett.106.077002, PhysRevB.84.024501, PhysRevB.84.064517, PhysRevLett.108.230509, PhysRevB.86.184514, Riste:2013vg, Wang:2014vm, PhysRevB.96.220501, PhysRevLett.121.157701, PhysRevApplied.12.014052, PhysRevB.100.140503, Vepsalainen:2020uj, Marin-Suarez:2020tf}, which are broken Cooper pairs out of the superconducting condensate at low temperatures. When a quasiparticle tunnels across the Josephson junction (JJ), there is a possibility of exchanging energy with the qubit, leading to depolarization.  Additionally, the associated change in charge parity can induce a small shift in qubit frequency, producing pure dephasing~\cite{Riste:2013vg}.  Although the exact mechanism of non-equilibrium quasiparticle generation is an open question, studies suggest that external radiation ~\cite{Ryan, Corcoles_QP}, including stray infrared and optical photons~\cite{Barends_IR, Houzet19} that could be present in applications such as transduction~\cite{Oscar_Painter}, ionizing radiation from environmental radioactive materials~\cite{Vepsalainen:2020uj}, and cosmic rays~\cite{Cardani}, lead to a higher density of broken Cooper pairs (i.e. quasiparticles). In this paper, we will focus on how intrinsic features of two dimensional (2D) transmons such as  capacitor metallization, geometrical design or uniformity of oxide layer in qubit JJs affect the density and dynamics of non-equilibrium quasiparticles. We provide a novel method for probing the homogeneity of JJ barriers within the qubits by analyzing the temperature dependent behavior of quasiparticle tunneling (QPT). Finally, we will address the question of what limitation quasiparticles and the quality of the junction barrier impose on qubit performance and coherence. 

Our measurements are based on the change in charge parity which switches sign whenever a single quasiparticle tunnels across the junction. This process is detectable since the tunneling generates a small shift in the qubit transition frequency. We focus on superconducting transmons with sufficiently large energy splittings between the odd and even charge parity branches for either the first or second excited states. Such a charge dispersion generally calls for designing qubits with smaller $E_J$/$E_C$ ratio than usual transmons, where $E_J$ is the Josephson energy and $E_C$ is the charging energy~\cite{PhysRevA.76.042319}. For this study, the $E_J$/$E_C$ ratio ranged between 20 and 50 for various types of qubits. We employ an experimental scheme pioneered in Refs.~\cite{Riste:2013vg, PhysRevLett.121.157701} that uses a Ramsey pulse sequence to map the charge parity state to the transmon state and thus record a time sequence of the charge parity switches that displays quasiparticle tunneling events.  A Fourier transform of this time sequence reveals a characteristic Lorenzian power-spectral-density spectrum (PSD) whose characteristic frequency roll-off provides the mean QPT rate. The details of the measurements can be found in the Methods section.

\dograph{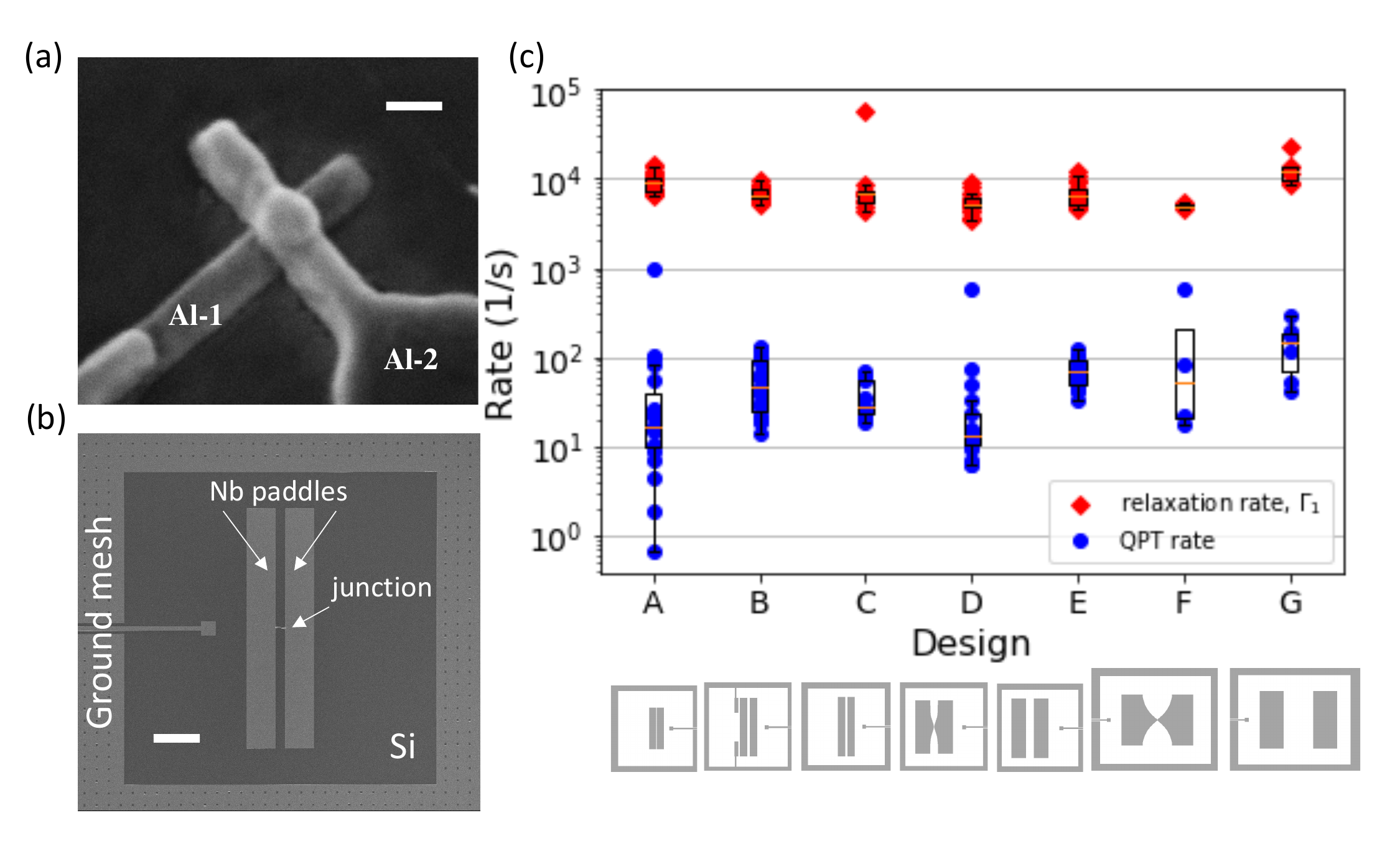}{width = 3.5 in}{\textbf{Non-equilibrium quasiparticles in various designs of Nb qubits.} (a) Scanning electron microscopy (SEM) image of a JJ in one of our transmon qubits showing the superconducting Al leads sandwiching a thin layer of AlO$_x$ layer. The scale bar corresponds to 100 nm. (b) SEM image of a standard transmon (design-C) incorporating a single Al/AlO$_x$/Al JJ and Nb capacior paddles. The scale bar corresponds to 100 $\mu$m. (c) QPT and relaxation rates vs. qubit design for a set of qubits with Nb paddles, the medians are shown in orange. The cartoons depict the various qubit designs used in this work from A to G with increasing paddle area.}

Our 2D transmon qubits are fabricated on high resisitivity Si substrates, where approximately 200-400 nm thick
metallization is sputter deposited and lithographically patterned using reactive ion etching to form large, coplanar capacitor
paddles~\cite{GambettaTAS17} from a variety of materials choices (see ``Methods'').  Single Al/AlO$_x$/Al junctions shown in Fig.~1a are formed using a standard Dolan bridge technique~\cite{Dolan77} and e-beam lithography. The shunting capacitor paddles are coupled to on-chip coplanar waveguide resonators enabling the readout and qubit control.  Although the standard superconductor used in such paddles is Nb, we have investigated qubits made of alternative superconducting materials, such as Ta, Al and NbN. Figure 1b shows a scanning electron microscopy (SEM) image of a standard transmon with Nb paddles that are separated by a 20 $\mu$m gap (design-C).  We have studied various transmons with different capacitor designs where we changed the distance between the paddles as well as the dimensions and the shape of the paddles; characteristic parameters of the qubits investigated in this study are summarized in Table 1. The devices in Fig.~1 were studied in the same packaging and measured in a light-tight enclosure in a cryogen-free dilution refrigerator with a base temperature of $\sim$12 mK. The measurements were conducted in completely identical fridge environments for all the devices presented throughout this paper.

\begin{table}[htp]
\caption{qubit parameters}
\begin{center}
\begin{tabular}{|c|c|c|c|}
\hline
Design & Style & capacitor gap  & paddle area  \\
      &   & ($\mu$m)      &  ($\mu$m$^2$)   \\
\hline
A        & non-tapered & 1.5& 300x60  \\
B        & non-tapered & 20 & 480x60 \\
C        & non-tapered &20 & 500x60   \\
D       &  tapered & 70 & 440x120 \\
E       &  non-tapered & 70 & 500x120 \\
F       & tapered &250&430x180 \\ 
G      & non-tapered & 250 & 480x200  \\

\hline
\end{tabular}
\end{center}
\label{table:deviceA}
\end{table}%

\section*{Results}
\subsection*{Non-equilibrium QPT rate at base temperature}

To study the impact of the non-equilibrium quasiparticles on coherence, we compare the energy relaxation rate
$\Gamma_1 \equiv$ 1/T$_1$ to QPT rate obtained at the base temperature. Figure~1c shows such data  for standard Nb qubits as a function of qubit design with increasing capacitor gap and paddle area. The QPT rates appear to be extremely slow among all the designs, making the parity switching times exceptionally long, ranging from 1 ms to up to 1.5 s. Thus, the qubits are not significantly disturbed by the QPT events during their average lifetimes. The observed QPT rates are substantially lower than those previously reported elsewhere, in which the parity switching times range between 
$\mu$s to ms~\cite{PhysRevLett.92.066802, PhysRevLett.97.106603, PhysRevB.73.172504, PhysRevB.77.100501, PhysRevB.78.024503, PhysRevLett.121.157701, Riste:2013vg, PhysRevB.100.140503}. The median of $\Gamma_1$s are at least two orders of magnitude larger than that of QPT rates suggesting that coherence of our standard devices is not currently limited by quasiparticle tunneling events. 

\subsection*{The role of materials in quasiparticle dynamics}

The microscopic properties of superconductors that vary from material to material are partially dictated by the charge dynamics, i.e. densities of paired (Cooper pairs) and unpaired charge carriers (quasiparticles). One such microscopic parameter is the kinetic inductance, which arises from the inertia of the charge carriers and is inversely proportional to the superfluid density of the film. The total kinetic inductance can reach significantly higher values for intrinsically low carrier density materials such as NbN compared to other superconductors with the same thickness ~\cite{Niepce, Annunziata} and have an impact on the dynamics of quasiparticles.

To determine how materials play a role in such dynamics and device performance, we explored alternative superconducting qubits where the capacitor paddles were made of Ta, Al and NbN and compared such devices with Nb qubits. Figure 2a and 2b show the comparison of QPT rate and qubit quality factor Q=2$\pi f_{01} T_1$ of alternative superconductors to those of Nb for a set of larger qubit designs, where $f_{01}$ is the qubit transition frequency.  One can see a discernible trend where Ta and Nb have comparable QPT rates, while NbN and Al have relatively larger values particularly in design-D. This trend reverses when the quality factor is plotted against the same designs for the same type of qubits; particularly the devices with Nb capacitors outperform those possessing Al and NbN capacitors in which the intrinsic QPT rate is greater.  For a given qubit design, despite having the larger superconducting gap, NbN exhibits discernibly lower quality factor compared to Nb. 

It is known in the field of microwave kinetic inductance detectors that quasiparticle loss is proportional to effective kinetic inductance fraction of a resonator~\cite{Zmuidzinas_review, Flanigan_thesis}.  However, for our transmon qubits, other sources of loss are thought to dominate the total quality factor and thus properties besides quasiparticle tunneling (such as surface loss) are likely the cause of the material-dependence of qubit quality factor \cite{Conal}.  Nonetheless, we can consider how kinetic inductance might influence QPT rates.

At temperatures close to absolute zero, the surface impedance of a superconductor is mainly reactive, where
the kinetic inductance scales with its London penetration depth~\cite{Zmuidzinas_review}.
The experimentally determined penetration depth of $\sim$300 nm within
400 nm thick, polycrystalline NbN films~\cite{Watanabe94},~\cite{Tang20} is much larger than
the value of 100 nm associated with 200 nm thick Nb films~\cite{Gubin05} and Ta~\cite{Greytak64}.
Paddle metallization with a larger London penetration depth may lead to a greater number of normal-state, conduction electrons, i.e: quasiparticles.  However, this trend is not consistent with the larger QPT rates observed in qubits posessing Al paddle metallization relative to those composed of Nb or Ta, as the penetration depth of Al is approximately
50 nm~\cite{Greytak64}.  We suspect its low superconducting gap energy may be responsible for a greater number of quasiparticles generated in the paddle metallization due to incident radiation.

\subsection*{Effects of qubit design on QPT}

In some of the devices, we observe a clear scaling in QPT with geometric parameters, such as qubit capacitor area. The QPT rates of NbN devices are shown as a function of design with increasing area in paddles in Fig.~3a. The same data are shown in Fig.~3b which exhibit a nonlinear trend as a function of paddle area.  Transmons with larger capacitor paddles show higher QPT rates, either due to the direct incidence of pair-breaking photons or phonon transmission of energy absorbed by the substrate.  Another interesting observation one can make is that the tapered designs help to suppress the QPT related dissipation. Note the reduced QPT rate in a tapered design D (F) with respect to a non-tapered design E (G), despite possessing similar paddle areas.  This suggests that the location of the paddles with respect to JJs is as important as paddle dimensions when determining the quasiparticle density in the devices~\cite{Gustavsson16}. In tapered designs, the distance between
the capacitor paddle sides and the JJ is reduced, possessing much shorter Al leads (see the inset of Fig.~3a).  This configuration may provide a more effective trapping mechanism for quasiparticles diffusing from the capacitor pads to the Al leads~\cite{Wang:2014vm}.  Furthermore, the paddles of non-tapered designs have long and narrow constrictions while such constrictions are absent in tapered designs.  These constrictions might contribute to larger kinetic inductance in non-tapered designs, leading to their higher QPT rates.  Finally, the lack of narrow constrictions in tapered designs could result in less current crowding, which we speculate could also lead to lower QPT rates in tapered designs.

\dograph{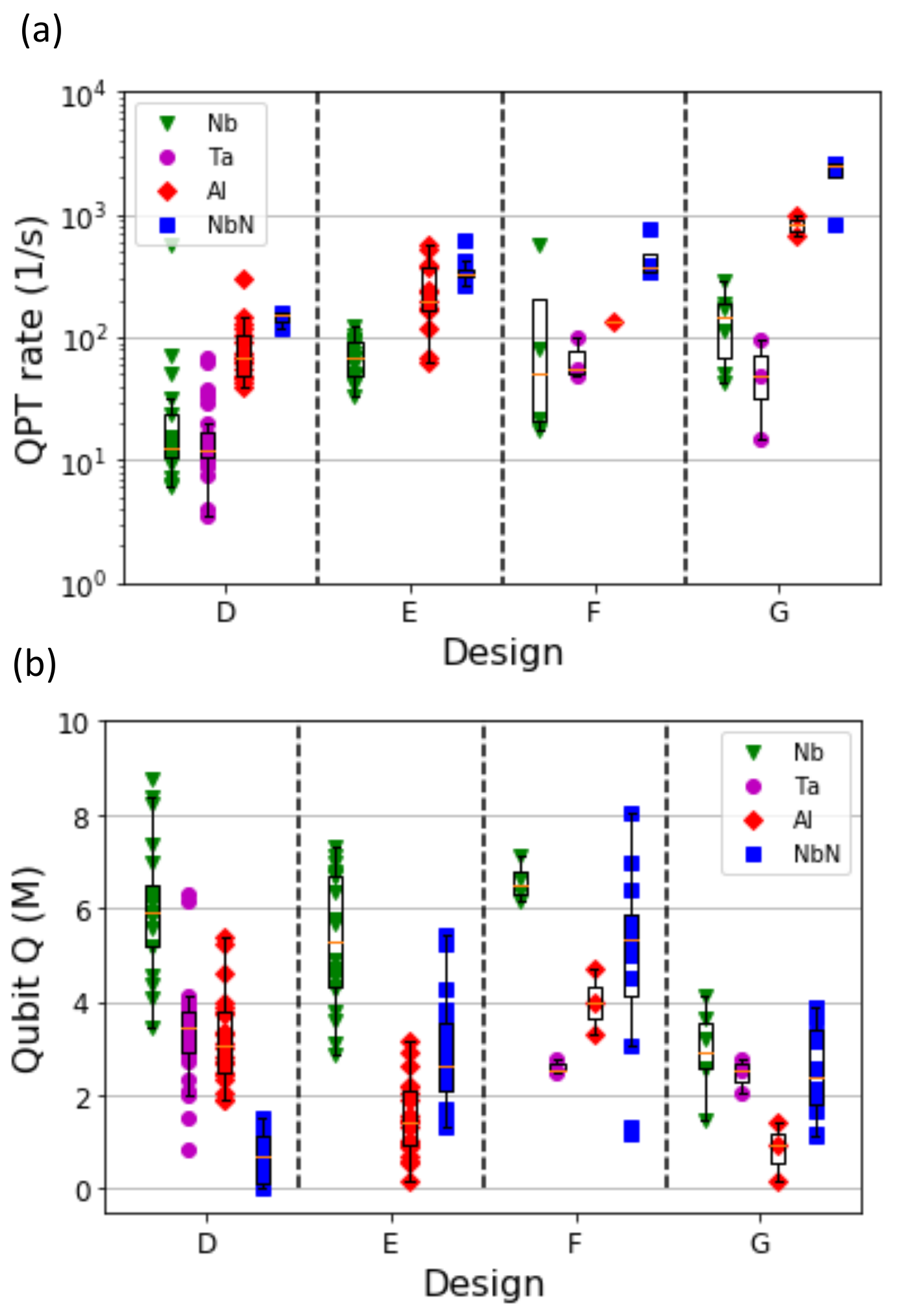}{width = 2.5 in}{\textbf{Behavior of qubits containing alternative superconductors.} Comparison of (a) quasiparticle tunnel rate and (b) quality factor Q=2$\pi f_{01} T_1$ of qubits with
various designs and capacitor paddles composed of different superconductor materials such as Nb, Ta, Al and NbN.}

To better understand how the design of the qubits can affect the quasiparticle generation and to determine the role of the tapering on QPT, we apply finite element method, electromagnetic simulations of our device geometries using HFSS 
(Ansys, Inc). In this model, we treat a single transmon qubit as an antenna by considering the 
reciprocity between radiation absorbed by this qubit and radiation emanating from the qubit 
in its excited state into an environment bounded by a surface that possesses a finite conductivity. While the HFSS simulations we conducted only involved microwave excitation, they are used as a proxy to assess the susceptibility of a given qubit design to incident energy (both directly and by transmission through the substrate). We calculate the real part of the admittance Re[Y($\omega$)] of the qubit
junction, which is proportional to the effective relaxation rate from this loss mechanism, Re[Y($\omega$)]/$C_q$, where $C_q$ is the total capacitance of the qubit~\cite{Houck2008} and $\omega$ is the excited to ground state transition frequency of the qubit~\cite{Rafferty2103}.
The magnitude of Re[Y($\omega$)], which is related to the current loss associated with the skin depth of the bounding box~\cite{Cheng83}, depends inversely on the square root of the conductivity of the qubit environment in the limit of low loss (see ``Supplementary Note 1"). We can therefore use a relative quantity, normalized by design-G possessing the largest paddles, to compare the different designs.  This metric follows an exponential trend with respect to qubit paddle area as shown in Fig.~3c and is expected to be linearly proportional to the relaxation rate~\cite{PhysRevB.67.094510, PhysRevB.84.064517}. Mean values of experimentally obtained QPT rates for NbN and Al qubits are plotted in Fig.~3d as a function of simulated Re[Y($\omega$)], confirming a linear dependence as the model dictates. However, the slopes of linear fits to the data from tapered and non-tapered designs are substantially different, reflecting the greater sensitivity that non-tapered capacitors 
possess to environmental radiation due to relatively larger paddle areas and more ineffective QP trapping. Note that the vertical axes of NbN and Al devices differ by factor of 2.8.


\dograph{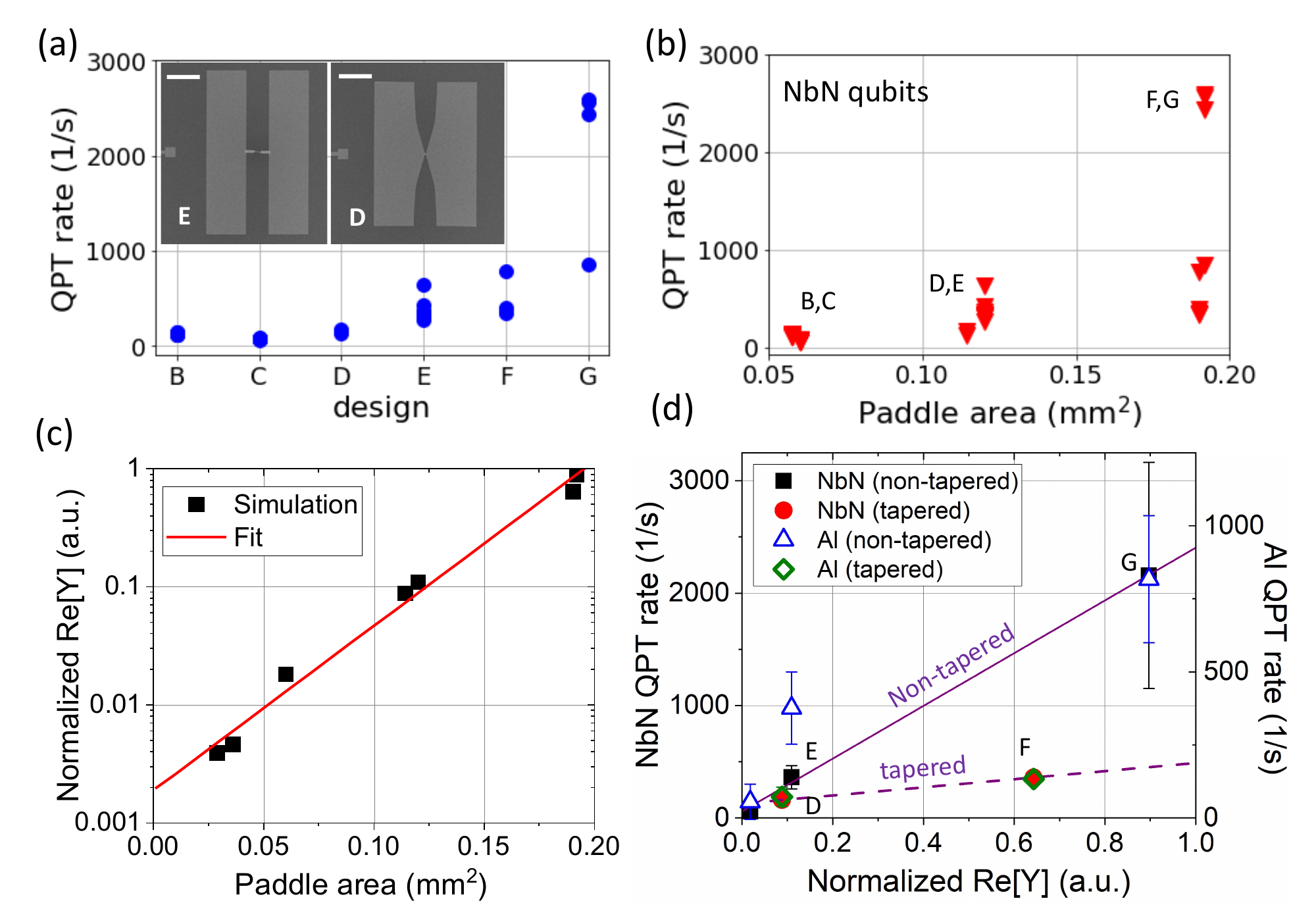}{width = 3.6 in}{\textbf{Impact of qubit design on quasiparticle dynamics.} (a) QPT rate vs. design in qubits with NbN paddles illustrating the nonlinear dependence of quasiparticle tunneling with increasing paddle dimensions, the inset shows  the SEM images of tapered (D) and non-tapered (E) qubit designs, the scale bars correspond to 100 $\mu$m. (b) QPT rate vs. paddle area for the same qubits, (c) real part of the simulated qubit admittance which follows an exponential increase with paddle area, d) comparison of mean values of experimentally obtained QPT rates for NbN and Al paddle qubits vs. simulated Re[Y($\omega$)].}

\subsection*{Temperature dependence of QPT rate}

Having established that both the material composition and geometrical design of the transmon paddles play an important role on QPT, we now turn our focus on the relation between the JJ barrier quality and QPT rates.  The QPT-induced relaxation rate scales linearly with quasiparticle density, which is an exponential function of the superconducting gap and temperature~\cite{deVisser11}. We utilize the temperature dependence of the QPT rate to infer the superconducting energy gap of Al and evaluate the oxide barrier uniformity in the JJs. Figure 4a shows such data from 10 qubits with NbN paddles on the same chiplet (design-E) sharing the same fabrication conditions. One can clearly observe two different temperature trends of QPT at low temperatures. Despite the variation in QPT rate at the base temperature, the blue curves display a behavior consistent with a distribution of conventional, non-equilibrium quasiparticles and an upturn in QPT with increasing temperatures signaling that thermal quasiparticles dictate the tunneling across the junction.  The red curves, however, demonstrate an unusual departure from the characteristic flat background of non-equilibrium quasiparticles at low temperatures before thermal quasiparticles dominate. We observed both types of temperature trends in a substantial number of qubits regardless of material or geometry of the capacitor paddles; this suggests that the JJ is the primary element (which is supposed to be identical in every type of qubit studied here) responsible for the anomalous characteristics rather than any other part of the qubit circuit. We have not found a direct correlation between the temperature profile of QPT and the qubit performance when the individual qubits (with the same capacitor material and geometry) are examined (see ``Supplementary Note 2").

If the JJ possesses an ideal tunnel barrier, i.e. the dielectric layer is homogenous and free of defects throughout the junction, the QPT rate can be modeled by~\cite{PhysRevB.84.064517}:
\begin{equation}\label{eq:qpt}
\Gamma_{qp} \sim \Gamma_{qp}^{ne}+ \sqrt{\frac{4\omega k_B T}{\hbar \pi}} e^{-\Delta_0/k_B T},
\end{equation}
where $\Gamma_{qp}$ exponentially scales with a single superconducting gap $\Delta_0$. The term $ \Gamma_{qp}^{ne}$ represents the non-equilibrium quasiparticles and corresponds to the flat background in the temperature sweeps. Now we consider a JJ with higher transmission regions, which can originate from defects in the junction barrier 
that might lead to spatially localized, quasiparticle-trap states or Al sub-gap states as detected previously in the grains of oxygen rich,
granular Al films~\cite{in_gap_states}.  We model the effects of such sites as possessing an effective energy gap
$\Delta_{i}$ less than that in the majority of the junction barrier corresponding to the Al gap $\Delta_0$.
The cartoon in the inset of Fig.~4a depicts the energy spectrum of effective energy gaps along the junction coexisting with the majority gap. The composite QPT rate can be written as a summation of parallel contributions associated with the majority gap and smaller effective energy gaps:
\begin{equation}
 \Gamma_{qp} = \Gamma_{qp}^{ne} + \sum\limits_{i}^n \Gamma_{qp}^{i}
 \end{equation}
For simplicity we consider only one additional channel representing the smallest effective gap, $\Delta_1$, which dominates
the other parallel loss mechanisms. The QPT rate can be written as (see ``Methods"):
\begin{equation}\label{eq:qpt1}
\Gamma_{qp} \sim \Gamma_{qp}^{ne}+ \sqrt{\frac{4\omega k_B T}{\hbar \pi}} \left[e^{-\Delta_0/k_B T} +A e^{-\Delta_1/k_B T}\right],
\end{equation}
 where A represents a collection of terms associated with the higher transmission path:
 \begin{equation}\label{eq:qpta}
 A=x_1 \frac{\pi}{\hbar \omega^2} \frac{\Delta_1}{R_1 C_q},
 \end{equation}
 $x_1$ refers to the relative fraction of quasiparticles tunneling through $\Delta_1$, its effective normal-state resistance
 $R_1$ and the total qubit capacitance $C_q$.
 
 \begin{figure*}
    \centering
    \includegraphics[width=5 in]{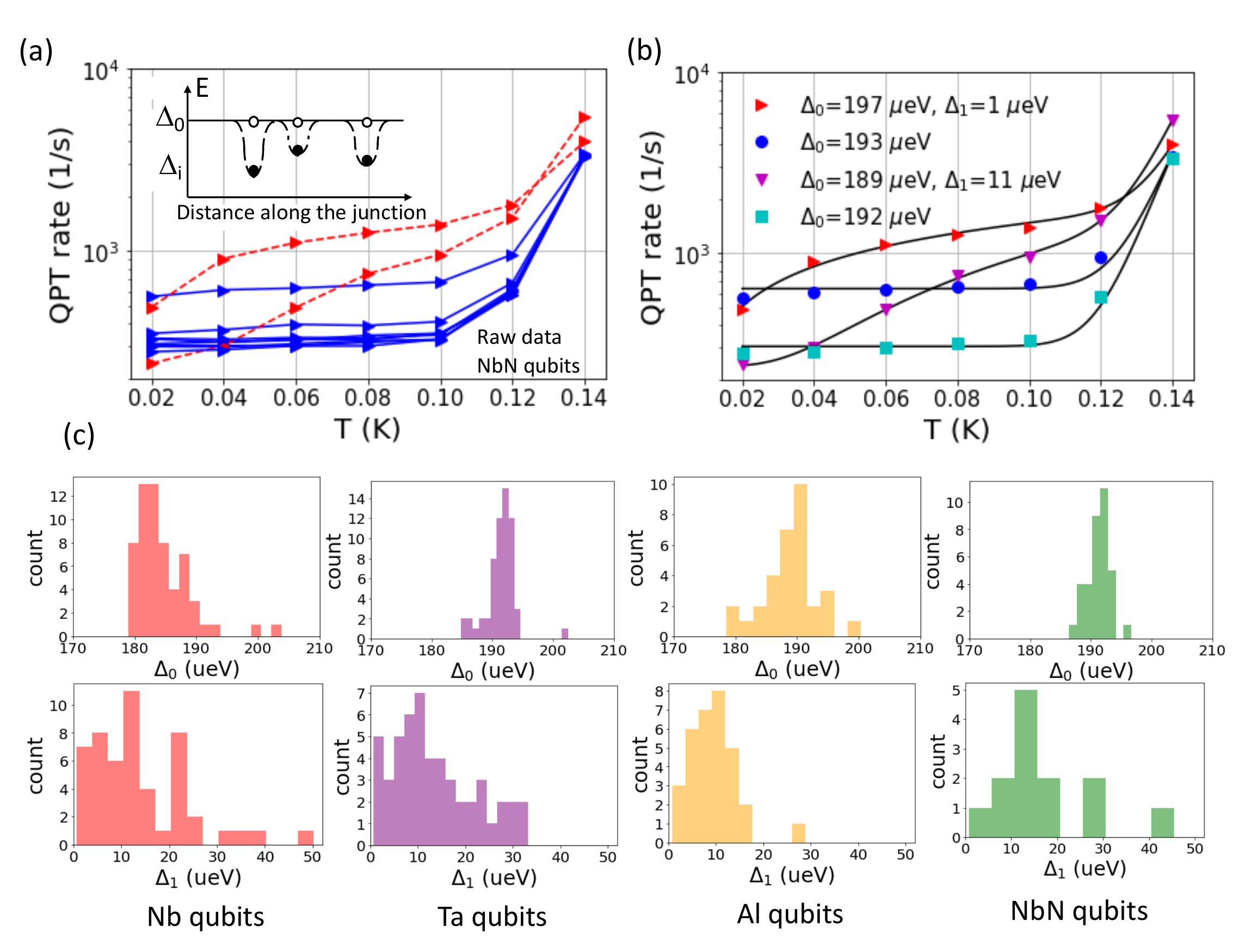}
    \caption{{\textbf{Temperature dependence of QPT rate.} a) Raw data of QPT rate vs. temperature for ten transmons with NbN capacitor paddles (with design-E). The functional form of the temperature dependence shows significant variation despite they share the same fabrication conditions. Note the difference between red and blue curves. The inset is a cartoon illustrating the energy spectrum of a junction where the barrier has high transmission sites. (b) Fits (black curves) to the QPT rate vs. temperature data (colored markers) showing a good agreement and consistency of the inferred Al superconducting gaps from both models. (c) The histograms built from fit parameters for various qubits with different type of paddle metallization.}}
    \label{fig:ds}
\end{figure*}

Figure~4b shows fits (black curves) to the data from four representative NbN qubits. Two of these curves have conventional behavior with a nearly constant non-equilibrium QPT rate up to $\sim$100 mK and the other two show strong deviations. The inferred superconducting gaps for Al, $\Delta_0$, are very similar regardless of the model described either in Eq.~(1) or Eq.~(3); however this is not the situation for the effective gaps, $\Delta_1$. We speculate that the extent of variation in $\Delta_1$ is due to the difference in barrier quality of the junctions in different qubits despite being fabricated in the same manner and having the same design. This variability is endemic to many instances of Al/AlO$_x$/Al Josephson junction fabrication, which is usually 
observed in the critical current~\cite{Kreikebaum20} or the corresponding junction frequency~\cite{Zhang20}, and
could result from subtle differences in the homogeneity of the angular deposition of Al, oxidation process, or photoresist quality across the wafers.  We argue that differences in the quasiparticle trap density or oxide thickness within the junction,
which have been proposed as responsible for critical current fluctuations and enhanced
conductivity~\cite{Rippard02, Greibe11}, may also correspond to these paths of increased QPT.

We have also collected the QPT rate vs. temperature data and analyzed it according to the models described above for qubits where the paddle metallization is Nb, Ta, or Al. Figure~4c shows the histograms formed from the extracted fit parameters for the Al superconducting energy gap $\Delta_0$ and the smaller effective gap $\Delta_1$. Each histogram includes results from 30$-$60 qubits with various capacitor styles and demonstrates a relatively small variation of $\Delta_0$, with median values of designs ranging between 183 and 193 $\mu$eV, which is consistent with literature values of the Al superconducting gap for similar films~\cite{PhysRevLett.106.077002, PhysRevLett.108.230509, Yamamoto}. The data forming the histograms of Nb paddle devices can be found in ``Supplementary Note 3" in the form of box plots showing the variation of the inferred parameters as a function of qubit design. We confirmed our results with independent cryogenic current-voltage measurements, which found a superconducting gap value of 185 $\mu$eV for slightly larger junctions with 200 nm Nb paddle metallization (see ``Supplementary Note 4"). However, we observe a broader distribution of $\Delta_1$, with median values of various designs ranging from 5 to 30 $\mu$eV, signaling a significant junction-to-junction variation.  A majority of the analyses shows that the effective gap associated with possible trap states or enhanced conduction in the junction is approximately 10$\%$  of the dominant Al gap (see ``Supplementary Note 3").


\section*{Discussion}

In conclusion, we have demonstrated a few orders of magnitude improvement on quasiparticle switching times over reported values, which translates into exceptionally low QPT rates. These direct measurements of switching times appear to be much longer than qubit lifetimes giving evidence that quasiparticles will not limit the coherence in the near future as we are trying to reach higher coherence times. We have found that the quasiparticle dynamics is intimately related to material type and geometry of the capacitors shunting the JJs. We have observed low temperature anomalies in the tunneling rate of non-equilibrium quasiparticles that are proposed to originate from defects or high transmission regions in the insulating barrier associated with either localized trap states or Al sub-gap states. Thus, careful analysis of temperature dependence of the QPT provides a valuable \emph{in situ} characterization of tunnel barriers within superconducting qubit junctions. 

\section*{Methods}

\subsection*{Qubit fabrication}

Transmon qubits employed in this study were fabricated using several different types of shunting capacitor metallizations.  The films were deposited on high resistivity Si wafers after native silicon oxide removal with a HF solution.  200nm Nb and 200nm Ta thin films were sputter deposited at room temperature, 400nm NbN was reactively sputtered with an Ar and N$_2$ gas mixture at 550C, and 200nm Al was deposited at 300C.  Nb, Ta, and NbN were subtractively etched with a Cl$_2$ based reactive ion etch using standard lithography.  Al/AlO$_x$/Al junctions are formed by shadow mask evaporation using e-beam lithographic patterning of PMMA/MMA resist in contact with the capacitor.  An ion mill was used immediately before the evaporation to remove the capacitor native oxide to improve contact.

\subsection*{Detection of charge-parity jumps}

The protocol developed in the pioneering work of Riste~\cite{Riste:2013vg} and
Serniak~\cite{PhysRevLett.121.157701, PhysRevApplied.12.014052} was adopted for this work with one significant
modification to facilitate QPT measurements on transmons with
$E_J / E_C \simeq 20$ or higher.  This method relies, as in previous
work, on mapping the charge-parity (CP) state of the transmon to the
transmon state using a Ramsey pulse sequence with delay time between
the Ramsey $\pi/2$ pulses chosen to give $\pm \pi / 4$ evolution in
phase with positive phase evolution for one CP state and negative for
the other.  The second Ramsey $\pi/2$ pulse is shifted 90$^{\circ}$ in
phase to map this phase evolution to the ground state population for
one CP state and the excited state for the other.  Measurement of the
transmon state before and after this Ramsey sequence gives the CP
state.  A rapid repetition of this sequence constitutes a time series
of samples of the CP state thus observations of QPT events
that switch the CP state.  This time series is Fourier transformed,
and the resulting power-spectral-density (PSD) of switching events fitted to
a Lorenzian function whose width yields the mean QPT rate.
Several such PSDs can be averaged to increase the SNR.  These
experiments are interleaved with simple Ramsey measurement of the CP
splitting which is used to set the correct delay time to achieve the
desired $\pi/4$ phase evolution.  Drifting of the CP splitting due to
changes in the overall charge environment of the transmon are detected
this way, and if significant drift occurs during a sampling sequence
(each typically lasting for a second or so), the data is discarded.

When $E_J / E_C$ is chosen to be large enough to minimize the charge
sensitivity of the transmon qubit, the charge dispersion of the 0-1
transition becomes small enough that the Ramsey delay time becomes
inconveniently long, reducing the CP state sampling rate and also
impacting the fidelity of the CP state mapping due to increased
probability of relaxation events.  To measure QPT in such
devices, we employ a modified pulse sequence.  Immediately following
the first measurement pulse, we transfer the populations of the 0 and
1 transmon states to the 1 and 2 states, respecively, using $\pi$
pulses on first the 1-2 transition and then the 0-1 transition.  The
CP-mapping Ramsey sequence is then executed on the 1-2 transition,
where the charge dispersion is much larger, facilitating a much
shorter delay time.  Populations of 1 and 2 states are then
transferred back to the 0 and 1 states by two more $\pi$ pulses before
the second measurement pulse.  The rest of the protocol proceeds as
before, with Fourier transformation and fitting of a Lorenzian
function to the PSD.  In transmons with intermediate values of
$E_J /E_C$, QPT rates can be determined using both the 0-1 and 1-2
transitions and compared.  We consistently observe excellent agreement
between the two methods, and generally higher overall CP mapping
fidelity using the 1-2 transition.

The CP mapping technique relies on setting the rf carrier frequency
midway between the charge-parity split transmon transition
frequencies.  We achieve this using a Ramsey pulse sequence with
linearly increasing phase of the second Ramsey $\pi / 2$ pulse to
mimic an effective frequency offset.  This allows the observation of
the beat frequency between the two transition frequencies as well as
the residual offset between the rf carrier and the average of the two
frequencies, and permits accurate adjustment of the frequency.

\subsection*{QPT model}

Following the work of Catelani~\cite{PhysRevB.84.064517}, we consider quasiparticle tunneling across the Josepshon
junction in a transmon qubit to be composed of parallel paths, including channels of high transmission probability~\cite{Catelani12}.  In the case of a homogeneous junction, the total quasiparticle tunneling rate,
$\Gamma_{tot}$ consists of both non-equilibrium,  $\Gamma_{ne}$ and thermal quasiparticle contributions, where the latter can be
related to the real part of the qubit admittance, Re[Y], and the qubit capacitance, C$_q$:
\begin{equation}\label{eq:gamtot}
\Gamma_{tot} = \Gamma_{ne} + \frac{Re[Y]}{C_q}
\end{equation}
Let us assume that the superconducting gap of the junction leads, $\Delta_0$, is much larger than the thermal energy,
k$_B$T, so that the Fermi-Dirac distribution, which governs the quasiparticle number density, can be approximated by
exp(-E/k$_B$T), where E is the quasiparticle energy.  Re[Y] can be simplified to form:
\begin{equation}\label{eq:rey1}
Re[Y] \sim \frac{1}{R_n} \frac{2 \Delta_0}{\hbar \omega} \left[ e^\frac{\hbar \omega}{2 k_B T}+e^{-\frac{\hbar \omega}
{2 k_B T}} \right] K_0 \left( \frac{\hbar \omega}{2 k_B T} \right) e^{-\Delta_0/k_B T}
\end{equation}
where R$_n$ is the normal state resistance of the junction, $\omega$ is the qubit transition frequency and K$_0$ is the
complete elliptic integral of the first kind.  Through the use of the Ambegaokar-Baratoff relation~\cite{AB63}, we can 
express R$_n$ in terms of the gap and junction inductance, L$_J$:
\begin{equation}\label{eq:rncq}
\frac{1}{R_n C_q} = \frac{\hbar}{\pi \Delta_0 L_J C_q} = \frac{\hbar \omega^2}{\pi \Delta_0}
\end{equation}
where the last step simply reflects that the angular frequency is the square root of 1/(L$_J$ C$_q$).  Combining
Eq.'s \ref{eq:gamtot} to \ref{eq:rncq}, and assuming that the qubit transition energy, $\hbar \omega >> k_B T$, we
arrive at Eq. \ref{eq:qpt} in the main text:
\begin{equation}\label{eq:repqpt}
\Gamma_{qp} \sim \Gamma_{qp}^{ne}+ \sqrt{\frac{4\omega k_B T}{\hbar \pi}} e^{-\Delta_0/k_B T}
\end{equation}

If high transmission paths exist for quasiparticles to tunnel through the junction, Re[Y] will now consist of a summation
across the parallel loss channels.  For simplicity, we only consider the one possessing the smallest effective resistance,
R$_1$, which should dominate quasiparticle tunneling across the junction.  Let us use an effective energy gap,
$\Delta_1$, to describe this high transmission path~\cite{Golubov04} so that we can generate an additional term to
include in the expression for Re[Y] from Eq. \ref{eq:rey1}:

\begin{equation}\label{eq:reyadd}
x_1 \frac{1}{R_1} \frac{2 \Delta_1}{\hbar \omega} \left[ e^\frac{\hbar \omega}{2 k_B T}+e^{-\frac{\hbar \omega}
{2 k_B T}} \right] K_0 \left( \frac{\hbar \omega}{2 k_B T} \right) e^{-\Delta_1/k_B T}
\end{equation}
where $x_1$ refers to the relative fraction of quasiparticles that tunnel through this path and we again assume that
$\Delta_1 > k_B T$.  However, we cannot apply the Ambegakor-Baratoff relation to simplify R$_1$.  In the limit of
$\hbar \omega >> k_B T$, the resulting formula for the total QPT rate can be approximated as:
\begin{equation}\label{eq:qpttot}
\Gamma_{qp} \sim \Gamma_{qp}^{ne}+ \sqrt{\frac{4\omega k_B T}{\hbar \pi}}
\left[e^{-\Delta_0/k_B T} + x_1 \frac{\pi}{\hbar \omega^2} \frac{\Delta_1}{R_1 C_q} e^{-\Delta_1/k_B T} 
\right]
\end{equation}

as shown in Eq's \ref{eq:qpt1} and \ref{eq:qpta} in the main text.

\section*{Data Availability statement}

The experimental data presented in this manuscript are available from the
corresponding author upon reasonable request.

\section*{Acknowledgements} We thank Zlatko Minev, Ted Thorbeck, Santino Carnavale, Elbert Huang, Dan Rugar and Oliver Dial for helpful discussions. We acknowledge John Ott for characterization of devices.  

\section*{Author Contributions}

C.K. performed the coherence $\&$ QPT measurements, analyzed the data and modeled the temperature dependence of QPT. C.E.M. performed the microwave simulations, developed the model for temperature dependence of QPT and guided the experiments. R.T.G. participated in the measurements and QPT analyses. B.B.W. designed and fabricated the qubits with alternative superconductors. Martin S. designed the earlier apparatus to perform QPT measurements and participated in the measurements. R.M.S. and A.E. developed the software performing the QPT measurements. V.A., A.A.S. and B.T. designed  $\&$ fabricated the qubits with Nb capacitors. A.F. assisted in modeling the QPT vs temperature data and measurements. E.R. performed the cryogenic I-V measurements. K.B. and J.S. developed the software for automating the measurements. A.P. participated in various stages of qubit fabrication. Matthias S. and K.R. advised on all efforts. All authors contributed to the discussion and production of the manuscript 

\section*{Competing Interests} The authors declare no competing interests.

\section*{Additional Information}  Supplementary Information.


\bibliography{QPT}

\vspace{200mm}

\end{document}